\documentclass[english,aps,twocolumn,prd,superscriptaddress,showpacs,preprintnumbers,amsmath,amssymb]{revtex4-1}

\usepackage{slashed}
\usepackage{lineno}
\usepackage{graphicx}
\usepackage{color}
\usepackage{cancel}
\usepackage{subfigure}
\usepackage{xcolor}
\usepackage{colortbl}
\usepackage{dsfont}

\definecolor{Gray1}{gray}{0.95}
\definecolor{Gray2}{gray}{1.}
\newcolumntype{a}{>{\columncolor{Gray1}[1.\tabcolsep]}c}
\newcolumntype{b}{>{\columncolor{Gray2}[1.\tabcolsep]}c}

\definecolor{columcol}{RGB}{0,153,76}
\newcolumntype{a}{>{\columncolor{columcol!20}[1.\tabcolsep]}c}
\newcolumntype{b}{>{\columncolor{columcol!7}[1.\tabcolsep]}c}
\newcolumntype{m}{>{\columncolor{columcol!50}[1.\tabcolsep]}l}
\usepackage{mathrsfs}
\definecolor{refcol}{RGB}{178,34,34}
\usepackage{dcolumn}
\usepackage{xspace}
\usepackage{bbold}
\usepackage{todonotes}
\presetkeys{todonotes}{color=green!20, size=\scriptsize}{}

\usepackage[colorlinks,linkcolor=refcol,citecolor=refcol,urlcolor=refcol]{hyperref}

\newcommand{\Tr}{\ensuremath{\operatorname{Tr}}}

\newcommand{\ua}{\ensuremath{U_A(1)}}

\def\eq#1{(\ref{#1})}

\def\Eq#1{Eq.~(\ref{#1})}
\def\Fig#1{Fig.~\ref{#1}}
\def\Sec#1{Sec.~\ref{#1}}
\def\App#1{App.~\ref{#1}}
\def\Tab#1{Tab.~\ref{#1}}

\definecolor{bjcol}{rgb}{0.12,0.56,1}
\definecolor{sircol}{rgb}{0.93,0.53,0.18}
\definecolor{fabcol}{rgb}{0.56,0.00,1.00}

\newcommand{\mm}{\marginpar{\colorbox{green}{\textbf{BJ}}\\@Mario:}}

\def\roughly#1{\mathrel{\raise.3ex\hbox{$#1$\kern-.75em%
\lower1ex\hbox{$\sim$}}}}

\newcommand{\diag}{\ensuremath{\operatorname{diag}}}

\def\ma0{m_{a_{0}}}
\def\mf0{m_{f_{0}}}

\newcommand{\ksb}{\ensuremath{k_{\chi\text{SB}}}}
\newcommand{\onefig}{0.90\linewidth}
\newcommand{\twofigs}{0.49\linewidth}

\graphicspath{
{./}
{./figures/}
}

\begin{document}

\title{Mass sensitivity  of the three-flavor chiral phase transition}

\author{Simon Resch} \email[E-Mail:
]{simon.resch@physik.uni-giessen.de} \affiliation{Institut f\"{u}r
  Theoretische Physik, Justus-Liebig-Universit\"{a}t Gie\ss en,
  D-35392 Gie\ss en, Germany\\[0.5ex]}

\author{Fabian Rennecke} \email[E-mail: ]{frennecke@quark.phy.bnl.gov}
\affiliation{Physics Department, Brookhaven National Laboratory,
  Upton, NY 11973, USA}

\author{Bernd-Jochen Schaefer} \email[E-Mail:
]{bernd-jochen.schaefer@theo.physik.uni-giessen.de}
\affiliation{Institut f\"{u}r Theoretische Physik,
  Justus-Liebig-Universit\"{a}t Gie\ss en, D-35392 Gie\ss en,
  Germany\\[0.5ex]}


\begin{abstract}
  The mass sensitivity of the chiral phase transition of QCD with and
  without axial $\ua$-symmetry breaking at vanishing and finite quark
  chemical potential is investigated. To focus on the low-energy
  sector of QCD, a quark-meson model with three dynamical quark
  flavors is employed.  Non-perturbative quantum fluctuations are
  taken into account with the functional renormalization group. The
  inherent ambiguities in fixing the low-energy model parameters away
  from the physical mass point and their consequences for spontaneous
  chiral symmetry breaking are discussed in detail and a heuristic
  parameter fixing scheme motivated by chiral perturbation theory is
  proposed. The influence of vacuum and thermal fluctuations of quarks
  and mesons on the order of the chiral phase transition is
  additionally assessed with a mean-field analysis.
\end{abstract}

\maketitle

\section{Introduction}

Quantum Chromodynamics (QCD) at finite temperature and density
predicts a phase transition at low energies from confined hadronic
matter to a deconfined quark-gluon plasma. The nature of the QCD
transition is subject of intense studies and relevant to ongoing and
planned heavy-ion experiments \cite{Gazdzicki:995681, BESwp,
  Friman:2011zz, Kekelidze2012, Galatyuk:2014vha, Sako:2014fha}.  The
global QCD symmetries associated to this transition are the chiral and
center symmetries, which are realized only in two antipodal extreme
limits of the quark masses. In the limit of infinitely heavy quarks,
i.e.~the pure gauge theory, the QCD vacuum obeys center symmetry. In
this limit a first-order transition to a phase with spontaneously
broken center symmetry occurs at a critical temperature of
$T_c \!\sim\! 270$ MeV \cite{Svetitsky:1982gs}.  In the opposite limit
of vanishing quark masses the QCD action is invariant under global
chiral symmetry and, at least for three massless flavors, also a
first-order transition is expected \cite{Pisarski:1983ms}.

For finite quark masses both chiral and center symmetry are explicitly
broken. As a consequence, the associated phase transitions weaken away
from both quark mass limits and finally terminate at critical points
of second-order transitions that belong to the three-dimensional $Z_2$
Ising universality class. For physical quark masses the chiral and
deconfinement transitions are smooth analytic crossovers with
approximately coinciding pseudocritical temperatures
\cite{Aoki:2006we, Borsanyi:2010bp, Bazavov:2011nk,
  Bhattacharya:2014ara}.

For the case of two degenerate light (up and down) quarks and one
strange quark the mass dependence of the order of the phase
transitions is summarized in the Columbia plot \cite{Brown:1990ev}. In
the heavy quark regime the Columbia plot, and even its extension to
finite baryon density, is by now well understood both from continuum
and lattice studies, e.g.~\cite{Fromm:2011qi, Saito:2013vja, Reinosa:2015oua,
  Fischer:2014vxa, Maelger:2018vow}. In the latter case, the sign problem becomes
treatable due to the large quark masses. The region of small masses is
less well understood, see e.g.~\cite{Karsch:2000kv, Karsch:2001nf,
  Karsch:2003va, deForcrand:2007rq, Varnhorst:2015lea, Jin:2014hea,
  Bazavov:2017xul, Jin:2017jjp} for lattice studies of the location of
the chiral critical line for three degenerate quark flavor. However,
the majority of lattice results confirm the existence of a first-order
region close to the three-flavor chiral limit. Although improved
calculations indicate a small first-order region, the results still
show huge discrepancies due to the notoriously difficult
implementation of chiral fermions on the lattice.

Another longstanding problem concerns the order of the chiral
transition in the chiral limit of two-flavor QCD. In the Columbia
plot, this is the point of massless light quarks and infinitely heavy
strange quarks. There is also a history of conflicting lattice results
in this case, e.g.~\cite{Iwasaki:1996ya, AliKhan:2000wou,
  Ejiri:2009ac, Bonati:2009yg}. The two-flavor case is particularly
intriguing since it is expected that the order of the phase transition
is most sensitive to the fate of the global $\ua$ symmetry
\cite{Pisarski:1983ms}. While this symmetry is anomalously broken in
the QCD vacuum \cite{'tHooft:1976up}, its restoration is expected at
large temperatures \cite{Gross:1980br}. If the restoration occurs
close to the two-flavor chiral phase transition in the chiral limit,
the transition can be of first order. Otherwise, a second-order
transition is expected and depending on the axial anomaly strength at
the chiral transition, the associated critical behavior can be
$SU(2)_L\times SU(2)_R \cong O(4)$ or $U(2)_L\times U(2)_R / U(2)_V$
whose corresponding critical exponents are very similar
\cite{Pisarski:1983ms, Aoki:2012yj, Pelissetto:2013hqa, Butti:2003nu,
  Grahl:2013pba}.  In case of a second-order transition in the
two-flavor and a first-order transition in the three-flavor chiral
limit, the $O(4)$ critical line of the light quark chiral limit and
the $Z_2$ critical line at finite quark masses meet at a tricritical
strange quark mass $m^{*}_s$. The existence and its relative location
of this tricritical point with respect to the physical mass point is
not established yet \cite{Li:2017aki} and even the possibility
$m_s^{*} \to \infty$ cannot be excluded \cite{Cuteri:2017zcb}.

An additional open problem is the interplay between the three- and
two-flavor chiral theories at finite chemical potential $\mu$. It is
not known whether a tricritical point $m^{*}_s$ at $\mu=0$, if it
exists, is analytically connected by a tricritical line to the
tricritical point at finite $\mu$ in the two-flavor theory. For a
positive curvature of the chiral critical surface, i.e.~when the
first-order region around the chiral limit increases with $\mu$, both
theories are connected analytically most probably and hence a critical
endpoint exists for physical quark masses.  However, if the size of
the first-order region shrinks for increasing $\mu$ the situation can
completely change.  Even if the two-flavor chiral theory exhibits a
tricritical point at some finite $\mu$ it need not be (analytically)
connected to the chiral critical surface which thus hampers any
conclusions about the existence of a critical endpoint in the phase
diagram \cite{Philipsen:2009dn}.

In summary, the nature of the chiral transition in the small mass
region of the Columbia plot is still controversial concerning several
lattice QCD studies for two and three quark flavors.  However,
functional continuum approaches such as the functional renormalization
group (FRG) or Dyson-Schwinger equations (DSE) do not suffer a sign
problem at finite density and the implementation of chiral fermions is
straightforward.  Renormalization-group methods such as the FRG are
also best suited to investigate universal features of phase
transitions.  Much progress towards an understanding of the QCD phase
structure from first principles has been made in recent years with
these methods, e.g.~\cite{Alkofer:2000wg, Fischer:2006ub,
  Bashir:2012fs, Fischer:2014ata, Mitter:2014wpa, Cyrol:2017ewj} and
references therein.  For example, FRG results predict a second-order
transition in the two-flavor chiral limit \cite{Braun:2009gm}.

Valuable qualitative and quantitative insights into the chiral phase
structure can be gained from effective low-energy QCD models such as
the linear sigma (LSM), quark-meson (QM) or Nambu--Jona- Lasinio (NJL)
models.  These models share the same global symmetries with QCD by
construction and exhibit similar or even the same symmetry breaking
pattern as the chiral transition in QCD.  Their criticality belongs to
the same universality class as QCD. Initiated by the first
renormalization group analysis of the LSM in \cite{Pisarski:1983ms},
the investigation of the Columbia plot with low-energy models has a
long history. For example, an investigation of the influence of the
axial $\ua$ anomaly on the chiral phase transition within the
Cornwall-Jackiw-Tomboulis (CJT) formalism in Hartree approximation of
the LSM yields a large first-order chiral transition region in absence
of the anomaly. Including the anomaly the results depend drastically
on the sigma meson mass and no first-order transition is found at all
in the most realistic case \cite{Lenaghan:2000kr}.  In
\cite{Herpay:2005yr, Herpay:2006vc} a zero temperature
parameterization for the mass dependence of the couplings based on
chiral perturbation theory ($\chi$PT) in an optimized perturbation
theory scheme for the LSM results in a first-order region around the
chiral limit which persists up to large strange quark masses. In
contrast, NJL models on a mean-field level predict a very small
first-order region \cite{Fukushima:2008wg}. In optimized perturbation
theory and the mean-field treatment of the QM model a large
first-order region in the chiral and in the light chiral limit is
obtained \cite{Kovacs:2006ym, Schaefer:2008ax}. Interestingly, in such
studies, the Columbia plot exhibits features which are qualitatively
independent of the axial anomaly \cite{Schaefer:2008hk}.

However, one drawback of these model investigations is the parameter
dependency which in parts can be reduced by combining effective
low-energy QCD models with the FRG such that an improved
non-perturbative truncation can be obtained in a systematical manner,
see e.g.~\cite{Gies:2002hq, Braun:2006jd, Braun:2014ata,
  Rennecke:2015eba}.

In this work, we go beyond these previous studies and perform a
non-perturbative FRG study of the Columbia plot in a three-flavor
quark-meson truncation at vanishing and finite density as well as with
and without the axial $\ua$ anomaly for the first time. The FRG allows
us to incorporate arbitrarily high loop orders as well as genuine
non-perturbative effects \cite{Berges:2000ew, Pawlowski:2005xe}. A
recent FRG study for a quark-meson truncation with $N_f= 2+1$
dynamical quark flavors has been done in \cite{Mitter:2013fxa} in a
leading-order derivative expansion, the local potential approximation
(LPA), where the effective potential is treated as energy scale
dependent, while all other correlation functions are kept constant. It
was found that the chiral transition in the light chiral limit at
physical strange quark masses is of second-order belonging to the
$O(4)$-universality class in the presence of the axial anomaly.
Without the anomaly, the transition was of first order in this
limit. The influence of scale dependent correlation functions beyond
the leading LPA has been worked out in
\cite{Rennecke:2016tkm}. However, the full Columbia plot has not been
studied with the FRG yet. Our approach allows us to systematically
study the impact of thermal and vacuum fluctuations of quarks and
mesons on the order of the chiral phase transition for arbitrary quark
masses.

Any investigation of the Columbia plot with a low-energy QCD
description faces the problem to fix the initial parameters
unambiguously. At the physical mass point the parameters can be
adjusted to reproduce certain experimental observables. Away from the
physical point no such phenomenological guidance can be found.  We
discuss the physical consequences of different parameter fixing
procedures and propose a heuristic fixing scheme motivated by chiral
perturbation theory ($\chi$PT).

This paper is organized as follows: after the introduction of our
effective low-energy description of QCD in terms of a quark-meson
model in \Sec{sec:model}, the incorporation of non-perturbative
quantum fluctuations by the FRG is summarized in the following
\Sec{sec:flucts}. In \Sec{sec:emfa} the vacuum fluctuations of quarks,
which are relevant for QCD thermodynamics, are included into the
conventional mean-field approximation.  Owing to the ambiguity in the
model parameters away from the physical point, two different parameter
fixing schemes are introduced in \Sec{sec:parafix}.  Our numerical
findings on the chiral critical lines in the Columbia plot at
vanishing and finite chemical potential with and without the axial
anomaly are presented in \Sec{sec:numerical-results}.  A summary with
an outlook can be found in \Sec{sec:sum}. Details of our numerical
implementation are given in the \App{sec:num}.

\section{Three-Flavor Quark-Meson Model}\label{sec:model}

In this section we briefly describe our setup of the three flavor
quark-meson model as a low-energy effective model for QCD. It is based
on a linear sigma model where the (pseudo)scalar meson nonets are
represented by a color-singlet $N_f \!\times\! N_f$ matrix field
$\Sigma$. It shares the same global chiral symmetry properties as QCD.
The inclusion of constituent quarks then facilitates the study of
finite chemical potential and significantly improves the chiral
dynamics such that, e.g., the chiral critical temperature at vanishing
density is much closer to lattice QCD, cf.~e.g.
\cite{Lenaghan:2000ey, Schaefer:2008hk}.  In general, quark-meson
models arise from QCD by successively integrating out quark and gluon
fluctuations towards low energies, see e.g., \cite{Braun:2008pi,
  Herbst:2010rf, Mitter:2014wpa, Braun:2014ata, Cyrol:2017ewj}.

Under the assumption that gluon fluctuations decouple below a certain
UV scale $\Lambda$, the $U_L(N_f)\otimes U_R(N_f)$ chiral invariant
part of the Euclidean effective action is given by
\begin{align}
  \label{eq:actionBosonized3}
  \begin{split}
    S_\chi[q,\bar q,\sigma_a,\pi_a]& = \int_x\Bigl\{\bar q
    \left(\slashed\partial + \gamma_0\hat\mu+ h\Sigma_5\right)q\\
    &\quad+\Tr(\partial_\mu\Sigma^\dagger\partial_\mu\Sigma)+U_{\chi}
    (\rho_1,\rho_2)\Bigr\}\,,
  \end{split}
\end{align}
with the three flavor quark field $q = (u,d,s)^T$
\cite{Jungnickel:1995fp, MeyerOrtmanns:1994nt}. The scalar and
pseudo-scalar meson fields $\sigma_a$ and $\pi_a$ are comprised in the
complex matrix fields
\begin{align}\label{eq:sigma}
  \Sigma=T^a(\sigma_a+i\pi_a)
\end{align}
and $\Sigma_5=T^a(\sigma_a+i\gamma_5\pi_a)$ where $T^a$ label the
$U(N_f)$ group generators. They are given explicitly
  by $T^0 \!=\! \mathds{1}_3 / \sqrt{6}$ and $T^i \!=\! \lambda_i/2$,
  with the eight Gell-Mann matrices $\lambda_i$, $i=1,\ldots,8$. The
action includes standard kinetic terms for the quark and meson fields
as well as a Yukawa type interaction vertex $\sim\!h$. A quark
chemical potential matrix $\hat\mu=\diag(\mu_u,\mu_d,\mu_s)$ is
included. In the following we restrict the discussion to symmetric
quark matter $\mu\equiv\mu_u=\mu_d=\mu_s$. We consider chirally
invariant meson-meson interactions of arbitrary order within an
effective chiral potential $U_{\chi} (\rho_1,\rho_2)$ which is a
function of the chiral invariants
$\rho_n = \Tr\left[(\Sigma^\dagger\Sigma)^n\right]$, $n\leq N_f$.
In the effectve potential a third invariant
$\rho_3$ has been dropped for the sake of simplicity.
If the chiral potential $U_{\chi}$ would only
depend on one invariant $\rho_1$, it would exhibit an enhanced
$SO(18)$ flavor symmetry. Adding a second invariant $\rho_2$
breaks this symmetry down to the desired chiral
$SU(3)_L\times SU(3)_R \times U(1)_A$ symmetry
\cite{Jungnickel:1995fp, Schaefer:2009ui}. Hence, the general
dependency on $\rho_1$ and $\rho_2$ represents a minimal
requirement to capture the correct flavor symmetry.

To effectively account for finite current quark masses as well as the
axial $\ua$ anomaly, we add the following symmetry breaking terms to the action:
\begin{align}\label{eq:Ssb}
  \begin{split}
    \mathcal{L}_\text{SB}(\sigma_a,\pi_a) &=
    -\Tr\left[j_aT^a(\Sigma^\dagger+\Sigma)\right] -c_A\,\xi\,,
  \end{split}
\end{align}
where the first term models the explicit breaking of the chiral
$SU_L(N_f)\otimes SU_R(N_f)$ due to finite quark masses.  The
anomalous breaking of the $\ua$-symmetry is described by the second
term which represents a bosonized version of the 't~Hooft determinant,
$\xi \!=\! \det(\Sigma^\dagger)+ \det(\Sigma)$ \cite{Kobayashi:1970ji,
  'tHooft:1976up, 'tHooft:1976fv}. The determinant gives a $N_f$-meson
interaction vertex which leaves
$SU_L(N_f)\otimes SU_R(N_f) \otimes U_V(1)$ intact but explicitly
breaks the $\ua$-symmetry down to $Z_A(N_f)$. Later, we will
effectively explore the effect of the anomalous breaking and consider
the two constant cases: $c_A=0$ and $c_A\neq 0$. However, in QCD this
coupling has also an explicit temperature and/or density as well as a
scale dependency which we ignore in the present work. For exploratory
studies see e.g. \cite{Pawlowski:1996ch, Gies:2002hq, Heller:2015box,
  Fejos:2016hbp, Fejos:2017kpq}.

The vacuum expectation values (VEV) for the fields are found at the
stationary point of the action. Due to parity and (approximate)
isospin symmetry as well as flavor neutrality of the vacuum only the
diagonal components of the scalar meson sector related to the
generators $T_0$ and $T_8$ can have finite expectation values.  Hence,
we assume light isospin symmetry and write $q = (l,l,s)^T$. At the
stationary point we can therefore express the chiral invariants and
the explicit symmetry breaking action, \Eq{eq:Ssb}, in terms of the
light and strange scalar meson VEVs
$\bar\sigma_l = \langle \sigma_l\rangle$ and
$\bar\sigma_s = \langle \sigma_s\rangle$ as follows:
\begin{align}
  \begin{split}
    \bar\rho_1                  &=\frac{1}{2} \left(\bar\sigma _l^2+\bar\sigma _s^2\right)\,,\\
    \bar\rho_2                  &=\frac{1}{8}\left(\bar\sigma _l^4+2 \bar\sigma _s^4\right)\,,\\
    \mathcal{L}_\text{SB}\big|_\text{EoM} &= -j_l \bar\sigma_l - j_s
    \bar\sigma_s - c_A\, \frac{\bar\sigma _s\bar\sigma
      _l^2}{2\sqrt{2}\,.\label{eq:UVEV} }
  \end{split}
\end{align}
$\bar\sigma_l$ and $\bar\sigma_l$ are directly related to the light
and strange chiral condensates $\langle \bar l l\rangle$ and
$\langle \bar s s\rangle$.  The fields denoted by the indices $l,s$
are expressed in the light-strange basis. They are related to the ones
in the singlet-octet basis, \Eq{eq:sigma}, via the rotation
\begin{align}\label{eq:ls08}
  \begin{pmatrix}\sigma_l \\ \sigma_s \end{pmatrix} =
  \frac{1}{\sqrt{3}}\begin{pmatrix} 1 & \sqrt{2} \\ -\sqrt{2} &
    1 \end{pmatrix} \begin{pmatrix}\sigma_8 \\
    \sigma_0 \end{pmatrix}\,. 
\end{align}
The pion and kaon decay constants can be expressed in terms of the
light and strange meson VEVs as \cite{Lenaghan:2000ey}:
\begin{align}\label{eq:decc}
  f_\pi = \bar{\sigma}_l\,,\qquad f_K = \frac{\bar{\sigma}_l+\sqrt{2}\bar{\sigma}_s}{2}\,,
\end{align}
which yield the quark masses
\begin{align}\label{eq:mcondrel}
  m_l=\frac{h}{2} \bar\sigma_l\,,\qquad m_s=\frac{h}{\sqrt{2}} \bar\sigma_s\,.
\end{align}
The current quark mass sensitivity of the chiral phase transition can
be controlled by varying the explicit symmetry breaking parameters
$j_l$ and $j_s$.
Since we have no direct access to the perturbative QCD regime within
the used low-energy description, it is more sensible to express the
Columbia plot in terms of the purely light pion- and the open strange
kaon masses, $m_\pi$ and $m_K$, instead of the corresponding current
quark masses. It is therefore useful to derive relations between the
explicit symmetry breaking sources and the pion and kaon masses,
\begin{align}\label{eq:jmassrel}
  j_l=f_\pi m_\pi^2,\qquad j_s=\sqrt{2}f_Km_K^2-\frac{1}{\sqrt{2}}f_\pi m_\pi^2\,.
\end{align}
The light-quark chiral limit $j_l = 0$ always implies $m_\pi = 0$,
while the strange quark chiral limit $j_s = 0$ leads to the relation
$m_K^2= f_\pi m_{\pi}^2/(2f_K) $. As a consequence, kaons are always
massive als long as the pions are. \Eq{eq:jmassrel} can be derived by
using the equations of motion
$\partial (U_\chi +\mathcal{L}_\text{SB})/\partial \sigma_{l,s} = 0$
together with the explicit expressions for $m_\pi^2$ and $m_K^2$ in App.\ D of
\cite{Rennecke:2016tkm}.

\section{Non-Perturbative Quantum Fluctuations}
\label{sec:flucts}

For an accurate description of the chiral phase transition all
quantum, thermal and density fluctuations, in particular in the
vicinity of the transition, should be taken into account in a
non-perturbative manner. As mentioned in the introduction this can be
achieved with the FRG, which is a semi-analytical method providing a
non-perturbative regularization and renormalization scheme for the
resummation of an infinite class of Feynman diagrams. For QCD related
FRG reviews, see e.g. \cite{Berges:2000ew, Pawlowski:2005xe,
  Gies:2006wv, Rosten:2010vm, Schaefer:2006sr, Braun:2011pp}.

\subsection{Functional Renormalization Group}
\label{sec:funct-renorm-group}

The central object of the FRG is a scale dependent effective action
$\Gamma_k$ that is identified with the classical action
$S = \Gamma_\Lambda$ at some UV-cutoff scale $\Lambda$. By
successively integrating out fluctuations momentum shell by momentum
shell, the full quantum effective action $\Gamma = \Gamma_{0}$,
i.e. the generating functional of all one-particle irreducible
correlation functions, is recovered in the infrared (IR) at
$k\rightarrow0$. The renormalization group flow of $\Gamma_k$ is
governed by the Wetterich equation \cite{Wetterich:1992yh}, which for
the quark-meson model reads
\begin{align}\label{eq:fleq}
  \begin{split}
    \partial_k\Gamma_k[q,\bar
    q,\sigma_a,\pi_a]&=\frac{1}{2}\text{STr}\left[\left(\Gamma_k^{(2)}+R_k\right)^{-1}_{\Sigma\Sigma} 
      \partial_kR_k^\Sigma\right]\\ 
    &\quad-\text{STr}\left[\left(\Gamma_k^{(2)}+R_k\right)^{-1}_{q\bar
        q} \partial_kR_k^q\right]\,.
  \end{split} \end{align} Here, $\Gamma^{(2)}_k$ denotes the second
functional derivative w.r.t. the corresponding meson or quark fields.
Fluctuations with momenta $p^2<k^2$ are suppressed by appropriate
regulator functions $R_k^{\Sigma,q}$. The super trace STr indicates
the trace over discrete (color, flavor, spinor) and continuous (loop
momenta) indices as well as various fields. The Wetterich equation
\eqref{eq:fleq} represents an exact functional equation whose solution
involves an infinite tower of partial differential equations. Hence,
in practice $\Gamma_k$ has to be truncated. In this work we choose the
local potential approximation (LPA) which amounts to a scale
dependence of the meson potential $U_{\chi, k}(\rho_1,\rho_2)$.
All other parameters are $k$-independent. Thus, the
scale-dependent effective action reads \begin{align}
  \Gamma_{k}[q,\bar q,\sigma_a,\pi_a]=S_{\chi,k}[q,\bar q,\sigma_a,\pi_a]+\!\int_x\!\!\mathcal{L}_\text{SB}(\sigma_a,\pi_a)\,,
  \label{eq:action}
\end{align}
where $S_{\chi,k}$ is given by \Eq{eq:actionBosonized3} with a
scale-dependent effective potential $U_{\chi}\rightarrow U_{\chi,
  k}$. The symmetry breaking part $\mathcal{L}_\text{SB}$, \Eq{eq:Ssb}, is
unchanged. In this approximation, and by employing optimized
regulators for the spatial momenta \cite{Litim:2000ci, Litim:2001up,
  Litim:2006ag}, the Wetterich equation turns into the following
partial differential equation for the effective potential:
\begin{align}
  \begin{split}
    \label{eq:flowEquation}
    \partial_k  U_{\chi,k} = \frac{k^4}{12\pi^2}
    &\Bigg\{\sum_{b=1}^{2N_f^2}\frac{1}{E_b} \coth\left(\frac{E_{b}}{2T}\right)\\
    &-2N_c\sum_{f=1}^{N_f}\frac{1}{E_{f}}\bigg[\tanh\bigg(\frac{E_{f}+\mu}{2T}\bigg)\\ 
    &\qquad \qquad
    +\tanh\bigg(\frac{E_{f}-\mu}{2T}\bigg)\bigg]\Bigg\}\,.
  \end{split}
\end{align}
The quark and meson quasi particle energies are abbreviated as
$E_i \!=\! \sqrt{k^2+m_i^2}$ with the corresponding light quark and strange quark masses,
$m_l \!=\! h\,\sigma_l/2$ and $m_s \!=\! h\,\sigma_s/\sqrt{2}$. The meson (curvature) masses $m_{b}$ are
obtained from diagonalizing the Hessian of the full effective
potential $U_{\chi,k}+\mathcal{L}_\text{SB}$, for explicit expressions we refer
to the appendix of Ref.~\cite{Rennecke:2016tkm}. Details on the
truncation and regularization of the effective action can be found in
Ref.~\cite{Mitter:2013fxa}. We remark  that improved
  LPA truncations  as well as  a different regulator choice  are
  likely to have some  quantitative effects on the presented
  results. In the present context the former has been studied  in
  \cite{Pawlowski:2014zaa, Rennecke:2016tkm} and the latter has been
  discussed in \cite{Pawlowski:2015mlf}. However, the qualitative
  statements in this work are expected to be unaffected by truncation
  and regularization effects.

\begin{table*}
  \renewcommand{\arraystretch}{1.2}
  \begin{tabular}{ c | c c c c | c c }
    \hline \hline
    &$a_{10,\Lambda} [\text{MeV}^2]$ & $a_{20,\Lambda}$ & $a_{01,\Lambda}$ & $c_A$ [MeV] & $m_\sigma$ [MeV] & $
\sqrt{m^2_{\eta'}+m^2_\eta}$ [MeV]\\
    \hline
    FRG & $562.02^2$ & $26.31$ & $50.38$ & $4807.84$ & 530 & 1103.6\\ 
    FRG & $345.84^2$ & $14.73$ & $108.70$ & 0 & 500 & 649.6\\ 
    \hline
    eMFA & $756.78^2$ & $11.40$ & $27.11$ & $4807.84$ & 530 & 1103.6\\ 
    eMFA & $666.07^2$ & $-4.21$ & $63.09$ & 0 & 500 & 649.6\\ 
    \hline \hline
  \end{tabular} 
  \caption{Initial UV parameters and the resulting experimental IR
    values for the FRG and eMFA calculation at the physical point.  The remaining parameters
    for the physical point are: $\Lambda=700$ MeV,
    $j_{l}^{(\text{phys})}=120.73^3\ \text{MeV}^3$, $j_{s}^{(\text{phys})}=336.41^3\ \text{MeV}^3$,
    $h=6.5$ with the corresponding IR observables $f_\pi=92.4$ MeV,
    $f_K=113$ MeV, $m_\pi=138$ MeV, $m_K=496$ MeV, and \mbox{$m_{l}=300$
    MeV}.}\label{tab:Parameters}
\end{table*}

\subsection{Extended Mean-Field Approximation}
\label{sec:emfa}

In order to understand the effect of the mesonic fluctuations we also
solve the quark-meson model in a mean-field approximation, which can
conveniently be achieved by switching off the meson contributions to
the flow equation \eq{eq:flowEquation}. The mesonic quantum fields are
replaced by their vacuum expectation values on the level of the
action. The remaining contribution is quadratic in the quark fields
and thus yields the conventional Gaussian path integral for the quark
contribution to the effective potential.

The quark contribution contains an UV-divergent vacuum part which
requires regularization. In standard (or no-sea) mean-field
approximation (MFA) this contribution is simply ignored.  However, as
shown in \cite{Skokov:2010sf} the vacuum contribution has a
significant impact on the chiral dynamics of the QM model.  In the
present context, the order of the chiral transition in the light
chiral limit depends crucially on fermionic vacuum fluctuations
\cite{Schaefer:2011ex}. For these reasons, we also explore the
Columbia plot in the extended mean-field approximation (eMFA), which
takes these vacuum fluctuations into account.

We want to emphasize that the vacuum contribution is automatically
included, and properly regularized, in the FRG approach. This also
includes the mesonic vacuum fluctuations. Hence, by switching off the
meson loops in \Eq{eq:flowEquation} we directly work in the eMFA.
Alternatively, the vacuum contribution of the QM model in eMFA can
also be regularized with standard techniques like e.g. dimensional
regularization yielding a renormalized QM model potential,
e.g.~\cite{Schaefer:2011ex}. However, performing a MFA within the FRG
it is guaranteed that the same non-perturbative regularization and
renormalization schemes are employed.

\subsection{Parameter Fixing}\label{sec:parafix}

Here, we discuss the initial effective action we use to initiate the
RG flow of our model. The general idea is to fix the couplings of the
effective action $\Gamma_{k=\Lambda}$, \Eq{eq:action}, at a UV scale
$\Lambda$ such that the resulting effective action $\Gamma_{k=0}$
reproduces phenomenologically known quantities in the infrared, like
meson masses and decay constants. The RG evolution is determined by
the flow equation \Eq{eq:fleq}.  For initial scales $\Lambda$ larger
than the scale of chiral symmetry breaking (in our case around
$\ksb \approx 500$ MeV) the VEVs of the mesons are small.  The meson
masses are large and consequently their fluctuations are
negligible. Therefore it is reasonable to expand the initial effective
meson potential about small VEVs of the mesons and keep only the
leading terms. It turns out for $\Lambda \gtrsim 700$ MeV finite
initial values for marginal and relevant terms are
sufficient. Furthermore, choosing $\Lambda \gtrsim 700$ MeV also
guarantees that UV cutoff artifacts at the relevant temperature and
chemical potentials are negligible. For a recent discussion regarding
UV cutoff effects in low-energy effective models we refer to
\cite{Braun:2018svj}. Irrelevant couplings, i.e.~those with a negative
mass dimension, are suppressed due to small initial meson
fluctuations. Their initial values can be set to zero. This yields the
initial potential
\begin{align}\label{eq:initU}
  \begin{split}
    U_\Lambda & = a_{10,\Lambda}\rho_1+
    \frac{a_{20,\Lambda}}{2}\rho_1^2+a_{01,\Lambda}\,\tilde\rho_2 \\ 
              & \quad-j_l\sigma_l-j_s\sigma_s-c_A\xi\,,
  \end{split}
\end{align}
where we have introduced a modified chiral invariant
$\tilde\rho_2=\rho_2-\rho_1^2/3$. For this ansatz seven parameters
$\left\lbrace a_{10,\Lambda} , a_{20,\Lambda}, a_{01,\Lambda}, c_A ,
  j_l, j_s, h\right\rbrace$ have to be fixed in the UV. At the
physical point, i.e.~for realistic masses, we can fix the initial
parameters such that we reproduce well-known masses and decay
constants in the IR. Note, that without anomalous breaking, $c_A=0$,
one has $m_{\eta^\prime} = m_\pi$ and $m_\eta$ is completely fixed by
the kaon and pion masses and their decay constants, cf.\
\cite{Lenaghan:2000ey}. For the mean-field approximation an identical
parameter fixing procedure with the same UV cutoff scale $\Lambda$ is
employed, see \cite{Schaefer:2008hk} for further details. The used
parameters are summarized in Tab.~\ref{tab:Parameters}.
  
We want to emphasize that irrelevant couplings, even though they are
zero initially, are generated during the RG flow towards the IR. This
has been demonstrated explicitly within a QM model in
\cite{Pawlowski:2014zaa}. In QCD, this follows from the fact that the
QCD effective action just above the chiral symmetry breaking scale is
dominated by effective four-quark interactions whose resonance signals
the spontaneous breaking of chiral symmetry \cite{Gies:2002hq,
  Mitter:2014wpa, Braun:2014ata, Cyrol:2017ewj}.
Based on this, we assume that higher order mesonic
  operators are generated only after the decoupling of the
  gluons. These operators are quantitatively and qualitatively
relevant for the non-universal physics considered here. We therefore
solve the flow equation for the full effective potential
$\partial_k U_{\chi,k}(\rho_1,\rho_2)$, \Eq{eq:flowEquation}, on a
two-dimensional grid instead of resorting to an expansion in terms of
only relevant and marginal operators, see App.~\ref{sec:num}.

In order to investigate the mass sensitivity and the influence of the
axial anomaly on the chiral phase structure the corresponding symmetry
breaking parameters $j_{l,s}$ and $c_A$ are varied and thus the system
is tuned away from the physical point. We cannot rely on IR
observables to fix the initial conditions in this case.  Since the
initial UV parameters are fixed in general by the underlying
fundamental theory of the strong interactions, it is \textit{a}
\textit{priori} not clear how the remaining parameters of the low-energy
model change when the current quark masses and/or the 't~Hooft
coupling are tuned away from the physical point.

For the axial symmetry parameter $c_A$, two scenarios are considered
(see Tab.~\ref{tab:Parameters}): one is where a constant $c_A$ is
chosen such that the $\eta^\prime$-meson mass is fixed to its
experimental value at the physical point. In the other case we assume
that the axial anomaly is absent, i.e.~$c_A=0$. The corresponding
initial UV parameters are fixed in such a way that the same
experimental IR values (as listed in \Tab{tab:Parameters}) are
obtained.  We want to emphasize that the latter case is not directly
connected to QCD, at least for small temperatures, since the QCD
vacuum always breaks the $\ua$-symmetry through quantum
fluctuations. Only for high temperatures \cite{Gross:1980br} or in the
large-$N_c$ limit \cite{Witten:1979vv} this symmetry is
restored. Thus, we implicitly assume the limit of infinite colors for
the $c_A=0$ computations.  Nonetheless, it is interesting to
investigate the influence of a $\ua$-symmetry on the phase structure
and study the consequences for the mass sensitivity from a more
general point of view. For exploratory FRG studies about
$\ua$-symmetry breaking, see \cite{Pawlowski:1996ch, Gies:2002hq,
  Heller:2015box, Fejos:2016hbp, Fejos:2017kpq}.

While varying the explicit chiral symmetry breaking $j_{l,s}$ is less
delicate, it gives rises to ambiguities for the parameter fixing of
low-energy models. In principle, the initial action of the low-energy
model at a given scale $\Lambda$ could be determined for varying
current quark masses by solving the corresponding QCD flow: Starting
with a microscopic QCD action at a perturbatively large energy scale
$k \gg 1$ GeV the RG evolution towards $\Lambda$ would yield the
desired initial conditions in an unique way. This is beyond the scope
of the present work, see, e.g., \cite{Springer:2016cji} for an
explicit example of a model parameter fixing
  procedure within an NJL model framework.
  
Owing to the scarce information about QCD away from
  the physical point, we explore two strategies to fix the parameters
  in this case:
\begin{itemize}
\item\emph{fixed-UV scheme:} Away from the physical point only the
  explicit symmetry breaking sources $j_{l,s}$ are assumed to change
  while all other parameters of the initial effective action are
  identical to the ones at the physical point. So the symmetric part
  of the initial effective action does not change in the UV.
\item\emph{fixed-$f_\pi$ scheme:} For each $j_{l,s}$ the initial
  effective action is adjusted such that the IR pion decay constant is
  always fixed to its physical value, i.e.  $f_\pi = 92.4$ MeV at
  every mass point in the Columbia plot.
\end{itemize}
The \emph{fixed-UV scheme} is practically the simplest choice and it
has been used frequently in the literature, e.g., \cite{Berges:1997eu,
  Lenaghan:2000kr, Fukushima:2008wg, Schaefer:2008hk,
  Herbst:2013ail}. In this scheme the initial action is given by the
parameters listed in Tab.~\ref{tab:Parameters} at the cutoff-scale
$\Lambda=700\,\text{MeV}$ and only $j_{l,s}$ are varied to explore the
quark-mass sensitivity of the phase transition. The underlying
assumption is that a change in the current quark masses of QCD can be
mapped directly onto a change of the symmetry breaking sources
$j_{l,s}$ in the effective low-energy model. The \emph{fixed-$f_\pi$
  scheme} is motivated by the findings of chiral perturbation theory
\cite{Gasser:1984gg}. In the light chiral limit, $j_l=0$, the pion
decay constant only slightly decreases. In the chiral limit when also
$j_s=0$, it does not change at leading order.  Since $\chi$PT suggests
a change of $f_\pi$ only on the order of 10\% towards the chiral
limit, we always adjust the initial conditions such to yield
$f_\pi = 92.4$ MeV in the IR, independent of $j_{l,s}$.  A more
accurate adaption to a more precise quark-mass dependence of $f_\pi$
(and also other observables) will only lead to minor quantitative
changes which are irrelevant in the present case. So in the
\emph{fixed-$f_\pi$ scheme} we demand (approximate) compliance with
$\chi$PT. A similar parameter fixing procedure for a linear
sigma-model based on $\chi$PT has been proposed in
\cite{Herpay:2005yr}.

\begin{table}[b]
  \renewcommand{\arraystretch}{1.3}
  \begin{tabular}{*{10}{c}}
\hline \hline
\multicolumn{5}{c|}{\emph{fixed-UV}}       &            \multicolumn{4}{c}{\emph{fixed-$f_\pi$}}        
\\
\hline
\multicolumn{1}{c|}{$\alpha$} & \multicolumn{1}{c}{$\Lambda$} & \multicolumn{1}{c}{$a_{10,\Lambda}$} & \multicolumn{1}{c}{$a_{20,\Lambda}$} & \multicolumn{1}{c|}{$a_{01,\Lambda}$} & 
\multicolumn{1}{c}{$\Lambda$} & \multicolumn{1}{c}{$a_{10,\Lambda}$} & \multicolumn{1}{c}{$a_{20,\Lambda}$} & \multicolumn{1}{c}{$a_{01,\Lambda}$}   
\\
\hline
\multicolumn{1}{c|}{$1$} & \multicolumn{1}{c}{$700$} & \multicolumn{1}{c}{$562.02^2$} & \multicolumn{1}{c}{$26.31$} & \multicolumn{1}{c|}{$50.38$} & 
\multicolumn{1}{c}{$700$} & \multicolumn{1}{c}{$562.02^2$} & \multicolumn{1}{c}{$26.31$} & \multicolumn{1}{c}{$50.38$}                                                                                        
\\
\multicolumn{1}{c|}{$0.17$} & \multicolumn{1}{c}{$700$} & \multicolumn{1}{c}{$562.02^2$} & \multicolumn{1}{c}{$26.31$} & \multicolumn{1}{c|}{$50.38$} & 
\multicolumn{1}{c}{$1000$} & \multicolumn{1}{c}{$562.02^2$} & \multicolumn{1}{c}{$26.31$} & \multicolumn{1}{c}{$50.38$}                                                                                        
\\
\multicolumn{1}{c|}{$0$} & \multicolumn{1}{c}{$700$} & \multicolumn{1}{c}{$562.02^2$} & \multicolumn{1}{c}{$26.31$} & \multicolumn{1}{c|}{$50.38$} & 
\multicolumn{1}{c}{$1143$} & \multicolumn{1}{c}{$562.02^2$} & \multicolumn{1}{c}{$26.31$} & \multicolumn{1}{c}{$50.38$}                                                                                        
\\
\hline \hline
\end{tabular}
\caption{Exemplary initial conditions for the two
  different parameter fixing schemes employed here (\emph{fixed-UV} and
  \emph{fixed-f$_{\pi}$}) for broken axial symmetry
  ($c_A \neq 0$). $\alpha$ parametrizes the value of the explicit symmetry 
  breaking sources $j_{l,s}$ according to \Eq{eq:alpha}. All
  units are the same as in \Tab{tab:Parameters}.}
  \label{tab:initcomp}
\end{table}

It is a priori possible that both schemes give the same
results. However, we will demonstrate that there are drastic
differences and discuss the physical reasons in the next section. The
Yukawa coupling $h$ and the anomaly coupling $c_A$ are not running in
our current approximation. In addition, we assume that they do not
change away from the physical point. The parameters that can
potentially be adjusted in the \emph{fixed-$f_\pi$ scheme} are
$a_{10,\Lambda}$, $a_{20,\Lambda}$, $a_{01,\Lambda}$, $ j_l$ and
$j_s$. Clearly, there is no unique choice for the initial parameters
in this case since we assume that $m_\pi,\,m_K$ and $f_\pi$ are the
only IR quantities we know away from the physical point. An exception
is the chiral limit, where also the relation between $f_\pi$ and $f_K$
as well as the masses of the pseudoscalar meson octet are fixed by
chiral symmetry. Our model consistently reproduces this case. It turns
out that for decreasing $j_{l,s}$ it is always possible to find a
larger initial scale $\Lambda^\prime > \Lambda$ with
$a_{10,\Lambda^\prime}=a_{10,\Lambda}$,
$a_{20,\Lambda^\prime}=a_{20,\Lambda}$ and
$a_{01,\Lambda^\prime}=a_{01,\Lambda}$ (cf.  \Tab{tab:Parameters})
such that the \emph{fixed-$f_\pi$ scheme} is realized. This procedure
is equivalent to a change in the initial parameters for a fixed
$\Lambda$. It is numerically much more convenient since everything is
controlled by only one parameter at given $j_{l,s}$. In
\Tab{tab:initcomp} we exemplify the initial conditions within the two
employed schemes for three different choices of explicit symmetry
breaking parameters $j_{l,s}$, parametrized as
\begin{align}\label{eq:alpha}
  \begin{pmatrix}j_l\\ j_s\end{pmatrix}=\alpha\begin{pmatrix}j_{l}^{\text{phys}}\\ j_{s}^{\text{phys}}\end{pmatrix}\,,
\end{align}
where $j_{l,s}^{\text{phys}}$ are the values at the physical point,
explicitly given in the caption of \Tab{tab:Parameters}.

\begin{figure*}
  \centering
  \subfigure[]{\includegraphics[width=0.35\textwidth]{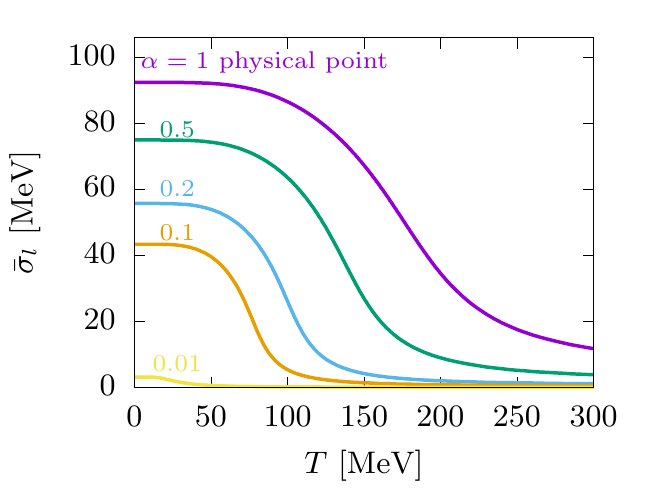}}\hspace{1cm}
  \subfigure[]{\includegraphics[width=0.35\textwidth]{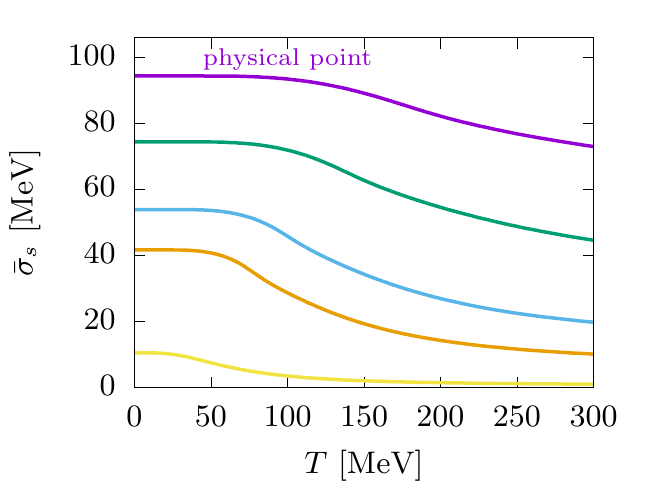}}
  \caption{Light and strange condensate as a function of temperature
    towards the chiral limit in the \textit{fixed-UV scheme}.  The
    parameter $\alpha$ linearly interpolates between the physical
    point ($\alpha=1$) and the chiral limit ($\alpha=0$), cf.\
    \Eq{eq:alpha}. Both condensates vanish in the chiral limit,
    signaling no spontaneous symmetry breaking (see text for
    details).}
\label{fig:chiralLimitVanishingCond}
\end{figure*}

\section{Numerical results}
\label{sec:numerical-results}

Lattice QCD calculations establish that for vanishing quark chemical
potential the chiral three-flavor phase transition is a smooth
crossover at the physical mass point \cite{Aoki:2006we}. This is also
a generic outcome of linear sigma- and quark-meson models.  In this
work we extend the investigation of the chiral phase transition for
arbitrary symmetry breaking sources $j_l$ and $j_s$ as well as finite
chemical potential.  We begin by reviewing various limiting cases:

\begin{itemize}
\item $j_s \rightarrow \infty$: Increasing $j_s$ above its physical
  value, the strange quark and all mesons with strangeness become
  heavier and eventually decouple from the flow towards the
  infrared. This limit is equivalent to the case of only two degenerate
  light flavors and, depending on the fate of the $\ua$ anomaly, will
  either resemble a
  $O(4)$-symmetric 
  or an $U(2)_L\otimes U(2)_R$-symmetric system.

\item $j_l = 0$ and $j_s > 0$: We call this limit the \textit{light
    chiral limit}.  It corresponds to two massless light
  quark flavors but finite strange quark masses and coincides with the
  ($m_\pi = 0$)-axis in the Columbia plot.

\item $j_l > 0$ and $j_s = 0$: This limit corresponds to the
  \textit{strange chiral limit}. The kaons then obey
  $m_K^2= f_\pi m_{\pi}^2/(2f_K)$. This defines the line of smallest
  possible kaon masses in the Columbia plot.

\item $j_l = j_s = 0$: All quark masses vanish and chiral
  $SU(3)_L\otimes SU(3)_R$-symmetry is not explicitly broken. Since
  this is the $N_f=3$ chiral limit, we refer to it as the
  \textit{chiral limit}.
\end{itemize}

\subsection{Spontaneous Symmetry Breaking}
\label{sec:ssb}

\begin{figure*} \centering
  \subfigure[]{\includegraphics[width=0.34\textwidth]{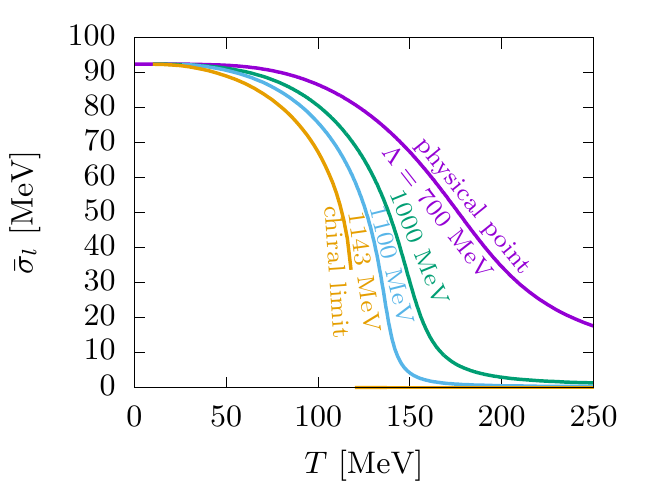}}\hspace{-.3cm}
  \subfigure[]{\includegraphics[width=0.34\textwidth]{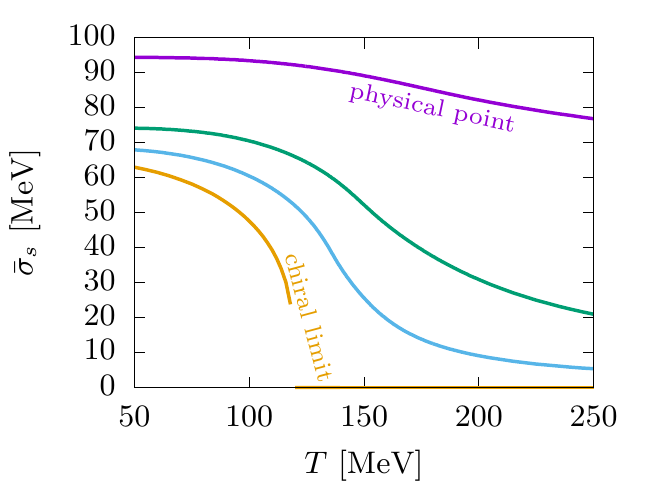}}\hspace{-.3cm}
  \subfigure[]{\includegraphics[width=0.34\textwidth]{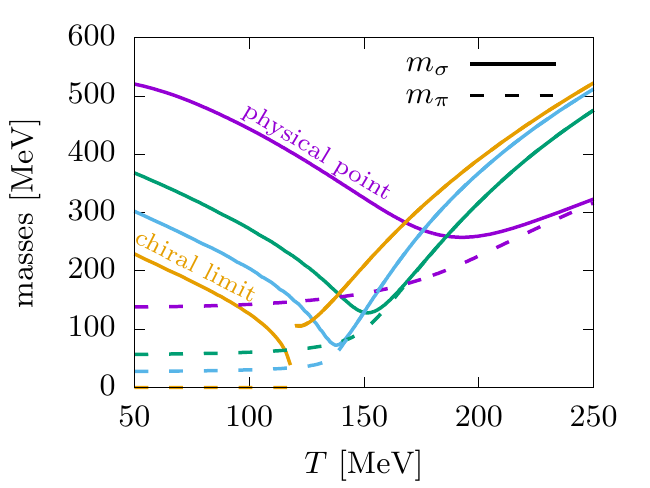}} \caption{Temperature
    dependence of the condensates (left and middle panel) and pion and
    sigma meson masses at $\mu=0$ towards the chiral limit in the
    \textit{fixed} \textit{$f_\pi$}-\textit{scheme}. For each
    $\Lambda$ the initial action has been adjusted such that
    $\bar\sigma_l=92.4$ MeV is fixed in the vacuum.
    $\Lambda=( 1143,\,1100,\,1000,\,700)$ MeV corresponds to
    $\alpha =(0,\,0.04,\,0.17,\,1)$. In the chiral limit a first-order
    phase transition is found.}
  \label{fig:chiralLimitCondensates}
\end{figure*}

As discussed in Sec.~\ref{sec:parafix}, the model parameter fixing
away from the physical point is not unique. In principle the effective
action $\Gamma_\Lambda$ at a given initial UV scale $\Lambda$ is
determined by an RG flow from the perturbative QCD regime down to
$\Lambda$. At least for the physical point this ambiguity in the
initial action of the low-energy model can be circumvented by
adjusting the initial action to known IR observables. As soon as the
physical point is left this procedure is not constructive
anymore.

In order to proceed we apply the in Sec.~\ref{sec:parafix} already
introduced two parameter fixing methods when the sources $j_{l,s}$ are
varied.  The \textit{fixed-UV scheme}, where all
other parameters remain fixed (cf.~\Tab{tab:Parameters}), has been
used in a FRG study of the light chiral limit \cite{Mitter:2013fxa}
and in a similar MFA study for the whole Columbia plot
\cite{Schaefer:2008hk}. In the latter case a rather heavy sigma mass
of the order of $800$ MeV was required in order to observe spontaneous
chiral symmetry breaking in the chiral limit, which is in conflict
with recent experimental data \cite{Patrignani:2016xqp}.

While the \textit{fixed-UV scheme} yields a reasonable value for
$f_\pi \approx 87 $ MeV in the light chiral limit which is in
agreement with chiral perturbation theory, we find that both light and
strange condensates vanish towards the chiral limit and spontaneous
chiral symmetry breaking is lost.  This is displayed in
Fig.~\ref{fig:chiralLimitVanishingCond} where the light (left panel)
and the strange (right panel) condensates are shown as a function of
the temperature for different points on a path connectiong the
physical point to the chiral limit in the Columbia plot. This path is
parametrized by a factor $\alpha \geq 0$ as defined in \Eq{eq:alpha},
such that $\alpha = 1$ corresponds to the physical mass point and
$\alpha = 0$ to the chiral limit.  Both condensates melt with
decreasing $\alpha$ and at the chiral limit they vanish. Furthermore,
the (pseudo)critical temperature $T_c$ drops, in particular
$T_c\rightarrow 0$ towards the chiral limit. The same behavior is seen
for the RG scale of spontaneous chiral symmetry breaking,
$\ksb$. Hence, the system remains in the symmetric phase in the chiral
limit.

This observation can be understood already on the mean-field level. In
this case the initial meson potential \Eq{eq:initU} matches the full
meson potential. For the light sector of the potential in the chiral
limit,
\begin{align}
  U_\text{MF}(\bar\sigma_l) = \frac{a_{10}}{2}\bar\sigma_l^2 +
  \frac{1}{8}\left(a_{20} + \frac{a_{01}}{3} \right) \bar\sigma_l^4\ ,
\end{align}
a finite expectation value $\bar\sigma_l$ can only be generated for
$a_{10}<0$ if the potential is bounded from below, which in turn
requires a positive quartic coupling $(a_{20} + a_{01}/3) > 0$.  With
the \textit{fixed-UV scheme} we find for sigma masses in the
range $m_\sigma\in[400,600]$ MeV a positive $a_{10}$ and thus
spontaneous symmetry breaking cannot occur. In the MFA this can be
circumvented by choosing unphysically large sigma masses
($m_\sigma\gtrsim 800$ MeV) \cite{Schaefer:2008hk}.

The above mean-field argument cannot directly be transferred to the
FRG case, since the full effective potential as a function of
arbitrary powers of the chiral invariants is evaluated beyond
mean-field. For instance, the FRG flow only allows for sigma masses in
the range $m_\sigma\in [400,600]$ MeV. Basically, $m_\sigma$ is
controlled by the quartic meson coupling. Vacuum stability of the
initial potential requires a positive quartic coupling which fixes the
lower bound for $m_\sigma$. On the other hand, the upper bound is
given by the requirement that mesons have to decouple at large energy
scales. Since the initial meson masses decrease with increasing
quartic coupling, this coupling cannot be arbitrarily large. In order
to understand why spontaneous chiral symmetry breaking is lost in the
\textit{fixed-UV scheme}, we need to understand the fluctuations that
eventually drive the system to criticality.

We observed that the values of the condensates, as well as $\ksb$ and
$T_c$, decrease with decreasing $\alpha$, i.e.~with less explicit
symmetry breaking. At first glance, this seems to be counter-intuitive
since smaller $j_{l,s}$ imply lighter current quark masses that yield
a potential enhancement of symmetry breaking fermionic
fluctuations. Furthermore, the quark masses are always lighter than
the RG scale above $\ksb$, i.e.  $m_{l/s,k} \ll k$ for
$k\in [\ksb , \Lambda]$, so that any variations in the current quark
masses should have less impact on $\ksb$. However, in the
\textit{fixed-UV scheme} the meson masses increase with decreasing
sources $j_{l,s}$ at the initial scale $\Lambda$. This also includes
the critical sigma mode which is related to the correlation length via
$\xi = 1/m_\sigma$.  Conversely, the initial correlation length
decreases with smaller $j_{l,s}$ such that (pseudo) criticality is
reached later in the RG flow. Hence, $\ksb$, the condensates and, as a
consequence, $T_c$ decrease with decreasing $j_{l,s}$.  The reason for
the sensitivity of the initial meson masses on the explicit symmetry
breaking parameters is that they change the evaluation point of the
effective potential that defines the physical parameters and not the
global form of potential itself, see, e.g.,
\cite{Pawlowski:2014zaa}. Thus, different $j_{l,s}$ can change the
meson masses substantially.

This observation contradicts our current understanding of chiral
dynamics in the chiral limit. For example, $\chi$PT predicts in the
chiral limit only minor modifications of the decay constants as
discussed in Sec.~\ref{sec:parafix}. In addition, this observation
might propose that the lower bound of the conformal window, $N_f^c$,
where spontaneous chiral symmetry breaking is lost for
$N_f \geq N_f^c$ chiral quarks in the fundamental representation,
would be $N_f^c=3$. However, functional continuum methods,
supersymmetry-inspired all-orders beta function approaches and lattice
QCD studies suggest $N_f^c \approx 8 - 12$, e.g., \cite{Braun:2006jd,
  Ryttov:2007cx, Appelquist:2007hu, Hasenfratz:2010fi, Bashir:2013zha}
and references therein. It is therefore reasonable to conclude that
the \textit{fixed-UV scheme} is not applicable within the present low-energy theory.
We note that this statement is unlikely to be affected by
truncation or regularization scheme effects within the QM model. The
main effect discussed above is the drastic increase of the initial
meson masses with decreasing $j_{l,s}$. It follows directly from
the form of the initial effective action. Only if the RG flow is
initiated at larger scales where gluon dynamics are relevant, our
results would be independent of the form of the initial meson
potential (as long as mesons are decoupled from the dynamics). This
clearly goes beyond the low-energy effective model employed here and
is discussed in detail in the context of dynamical hadronization in
\cite{Gies:2002hq, Mitter:2014wpa, Braun:2014ata}.

\begin{figure*}
  \subfigure[$\;c_A\neq
  0$]{\includegraphics[width=\twofigs]{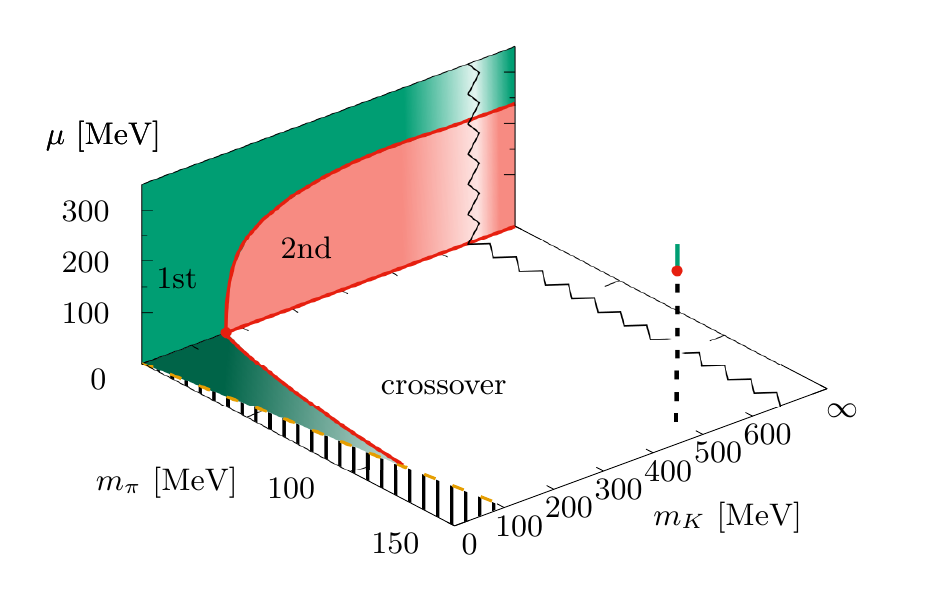}}
  \subfigure[$\;c_A=0$]{\includegraphics[width=\twofigs]{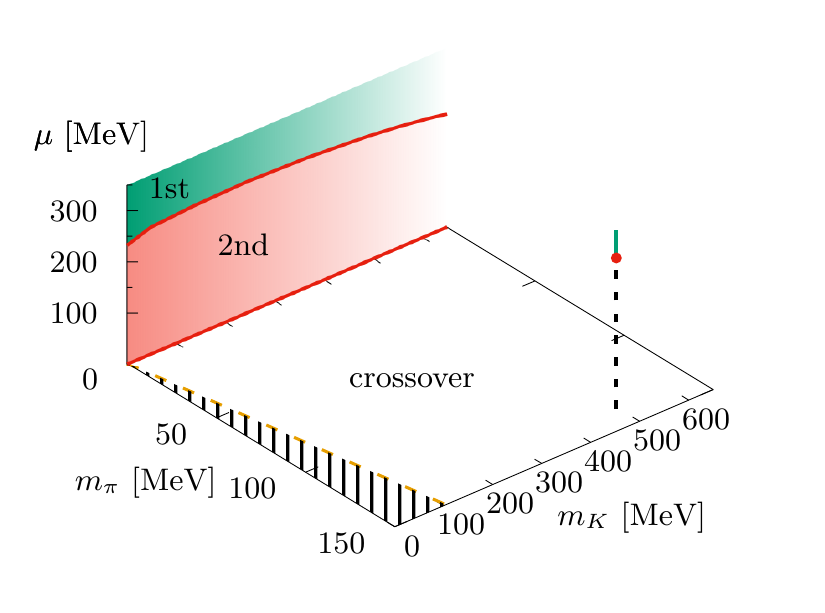}}
  \caption{ Chiral phase transition without meson fluctuations (eMFA)
    in the $(m_\pi , m_K)$- and $(\mu ,m_K)$-plane (left panel with
    $\ua$-symmetry breaking, $c_A\neq 0$, right panel without,
    $c_A=0$). First-order and crossover regions are separated by a
    second-order chiral critical line (solid red line). The vertical
    line indicates the physical mass point with a critical point (red
    point), crossover (black dashed line) and first-order transition
    (green solid line). In the dashed area, close to the $m_\pi$-axis,
    $j_s$ is negative. The orange dashed line corresponds to the
    strange chiral limit where $j_s = 0$.}
  \label{fig:columbiaPlotEMFA}
\end{figure*}

In the following we assume that chiral symmetry of the QCD vacuum is
spontaneously broken in the chiral limit. By demanding compliance with
$\chi$PT we introduced the \emph{fixed-$f_\pi$ scheme} in
Sec.~\ref{sec:parafix}. It follows from \Eq{eq:decc} and the symmetric
form of the initial effective potential \eq{eq:initU} that fixing
$f_\pi$ to a finite value for any $j_{l,s}$ always implies
spontaneously broken chiral symmetry in the infrared.  As discussed
above, if we fix $a_{10,\Lambda}$, $a_{20,\Lambda}$ and
$a_{01,\Lambda}$ at a given UV cutoff $\Lambda$, a decrease of the
explicit symmetry breaking parameters $j_{l,s}$ not only decreases the
pion and kaon masses in the IR but also leads to an unphysically large
reduction of the initial correlation length.  Since the quark
fluctuations that drive the system to (pseudo) criticality do not
change much for scales $k \!>\! \ksb$ and decreasing $j_{l,s}$, we
expect that the same behavior holds for the correlation length.  Due
to the intimate relation between the decay constants and the chiral
condensates, a larger UV correlation length also entails larger IR
decay constants. So the \emph{fixed-$f_\pi$ scheme} resolves the
problems of the \textit{fixed-UV scheme} by construction.  At the
physical mass point our initial conditions are fixed such that
$f_K/f_\pi \approx 1.2$, in agreement with lattice QCD
\cite{Aoki:2013ldr}. As we will demonstrate below, our procedure
naturally leads to $f_K = f_\pi$ in the chiral limit, in accordance
with chiral symmetry. Thus, the \textit{fixed}
$f_\pi$-\textit{scheme}, backed up in particular by $\chi$PT, seems to
be a physically reasonable procedure to fix the initial conditions
away from the physical mass point as it provides a natural
interpolation between the physical point and the chiral limit. A more
accurate procedure to fix the initial conditions that also takes the
scaling, e.g., of $f_\pi$, $f_K$, $m_\pi$ and $m_K$ near the chiral
limit as predicted by $\chi$PT into account is beyond the scope of the
present work. But since the physical picture presented here is
expected to hold in general, we anticipate only minor quantitative
changes.

Our findings within the \textit{fixed} $f_{\pi}$-\textit{scheme} are
shown in \Fig{fig:chiralLimitCondensates} where the light and strange
condensates (left and middle panel) as well as the pion and sigma
meson masses (right panel) as a function of temperature for different
UV cutoffs towards the chiral limit from $\Lambda=700$ MeV at the
physical point to $\Lambda\sim 1143$ MeV in the chiral limit are
shown. Again, a straight path in the Columbia plot from the physical
point to the chiral limit has been chosen, see \Eq{eq:alpha}. The
light condensate $\bar{\sigma}_l$ is always fixed to $f_{\pi}= 92,4$
MeV at $T=0$ by construction for every $\alpha$. With decreasing
$\alpha$ the chiral phase transition becomes steeper and turns into a
first-order transition below a critical value $\alpha_c\lesssim
0.04$. At $\alpha_c$ (not shown in the figure) the sigma mass
drops to zero at the critical temperature signaling a second-order
phase transition. Since the sigma meson is the only critical mode at
$\alpha_c$, the transition lies in the three-dimensional Ising
universality class.

The constituent quark mass consists of the current quark mass and a
contribution from spontaneous chiral symmetry breaking proportional to
the chiral condensates. Thus, towards the chiral limit no significant
changes for the light quarks are expected, while the strange quark
constituent mass decreases substantially.  Since $\bar\sigma_s$ is
connected to the strange constituent quark mass via \Eq{eq:mcondrel},
it also decreases significantly with decreasing $\alpha$. In fact, in
the chiral limit chiral symmetry implies the relation
$\bar\sigma_l = \sqrt{2}\bar\sigma_s$, which results in $f_K = f_\pi$,
cf.~\Eq{eq:decc}. The pseudoscalar meson octet with the pions, kaons
and the $\eta$-meson becomes massless. The pseudoscalar singlet
$\eta^\prime$-meson remains massive in the presence of the $\ua$
anomaly. The masses of the scalar meson nonet with the sigma, $f_0$,
$a_0$ and the kappas, are generated solely by spontaneous chiral
symmetry breaking. This is exactly what we find.

\subsection{Columbia Plot in  Mean-Field Approximation}
\label{sec:extended-mean-field}

We begin with an investigation of the Columbia plot in the extended
mean-field approximation as outlined in Sec.~\ref{sec:emfa}.
We express the
symmetry breaking in terms of the pion and kaon masses, $m_\pi$ and
$m_K$, by using the identities in \Eq{eq:jmassrel}.  We will
concentrate on lower and moderate mass values in the Columbia plot
since the opposite quenched mass limit is unreachable without direct
access to gauge degrees of freedom.  As parameter fixing we also apply
the \textit{fixed} $f_\pi$-\textit{scheme} by adjusting the UV cutoff
$\Lambda$ such that $f_\pi$ is fixed in the vacuum for any values of
$m_\pi$ and $m_K$. This extends the analysis in
\cite{Schaefer:2008hk}, where the Columbia plot has been computed in
standard MFA where vacuum quark fluctuations are omitted.  Together
with the following FRG investigation, we are now able to distinguish
between effects stemming from fermionic (vacuum as well as thermal)
and/or mesonic fluctuations.

In standard MFA the chiral transition is of first order in the light
chiral limit, apparently independent of $m_K$ and of the presence of
the axial anomaly. Particularly, this would imply a first-order chiral
transition in the two-flavor limit.

In \Fig{fig:columbiaPlotEMFA} the mass sensitivity of the chiral phase
transition in the ($m_\pi, m_K$)-plane as well as for finite chemical
potentials in eMFA are shown.  The results differ significantly from
the MFA results in \cite{Schaefer:2008hk}. Including the anomalous
$\ua$-symmetry breaking (left panel) we find a distinctive first-order
region around the chiral limit for vanishing chemical potentials. The
transition is of second-order in the light chiral limit for kaon
masses larger than the physical ones and in particular in the two
flavor limit. These findings agree with the predictions in
\cite{Pisarski:1983ms} for the three and two flavor chiral
limits. Since standard MFA predicts a first-order transition
\cite{Skokov:2010sf}, our results imply that the second-order
transitions in the light and two-flavor chiral limits are induced by
fermionic vacuum fluctuations. Similarly, and in contrast to a
standard MFA study \cite{Schaefer:2008hk}, a tricritical point
$m_K^{*}= 169$ MeV on the $m_K$-axis for $\mu=0$ is found when quark
vacuum fluctuations are taken into account.  At this point the
second-order chiral transition line in the light chiral limit
terminates and the transition turns from second- to first-order with
decreasing $m_K$. In the opposite direction for finite pion masses the
chiral critical line terminates at the $j_s = 0$ axis (dashed orange
line in the figure) corresponding to $m_\pi \approx 110$
MeV. Everywhere else the transition is a smooth crossover for $\mu=0$.

For $\mu>0$ the size of the first-order region around the chiral limit
increases such that the chiral critical surface has a positive
curvature, cf.~\cite{deForcrand:2006pv}. Consequently, a chiral critical
endpoint exists in the ($T,\mu$)-phase diagram for physical
masses which is marked as a red point at the vertical line in
\Fig{fig:columbiaPlotEMFA} and represents the intersection point with
the chiral critical surface \cite{deForcrand:2007rq}. In this scenario
the found tricritical point $m_K^{*}$ at $\mu=0$ is expected to be
analytically connected by a tricritical line to the tricritical point
at some finite $\mu$ in the two flavor chiral limit by increasing
$m_K$ \cite{Halasz:1998qr}. This behavior is also confirmed in
\Fig{fig:columbiaPlotEMFA}: for increasing $\mu$ and larger $m_K$ the
chiral critical line has a positive slope and saturates for
$m_K \gtrsim 500$ MeV at $\mu_c=239$ MeV and is continuously connected
to the critical endpoint in the two-flavor chiral limit. For chemical
potential below the chiral critical line the chiral transition is
always of second order for $m_{\pi} =0$ and first-order above.

Without the $\ua$-symmetry breaking (right panel in
Fig.~\ref{fig:columbiaPlotEMFA}) no first-order phase transition for
$\mu=0$ is found at all. This is in disagreement to
\cite{Pisarski:1983ms} as well as standard MFA results
\cite{Schaefer:2008hk}. A second-order transition is found in the
light chiral limit independent of $m_K$ when the quark vacuum
fluctuations are included and a first-order one if not. Hence,
fermionic vacuum fluctuations induce a second-order transition in the
(light) chiral limit. A first order region is only seen at larger
chemical potentials. In the chiral limit at $\mu_c=232$ MeV which
rises to $\mu_c=256$ MeV for a kaon mass around $m_K=243$ MeV and then
slowly declines again. At the physical point, we find a critical
endpoint at $\mu_c \sim 298$ MeV.

Note that in the standard MFA analysis \cite{Schaefer:2008hk} an
extended first-order region along the $m_K$-axis is seen independent
of the $\ua$-symmetry breaking. However, in order to obtain a
spontaneous symmetry breaking in the chiral limit a large sigma meson
mass was chosen. We have also verified that with vacuum fluctuations
and a comparatively large sigma meson mass similar results as in
\Fig{fig:columbiaPlotEMFA} are obtained. This confirms that the
differences we found can indeed be attributed to fermionic vacuum
fluctuations.

\subsection{Columbia Plot with the FRG}
\label{sec:columbia-plot-with}

\begin{figure}[t]
  \centering
  \includegraphics[width=\onefig]{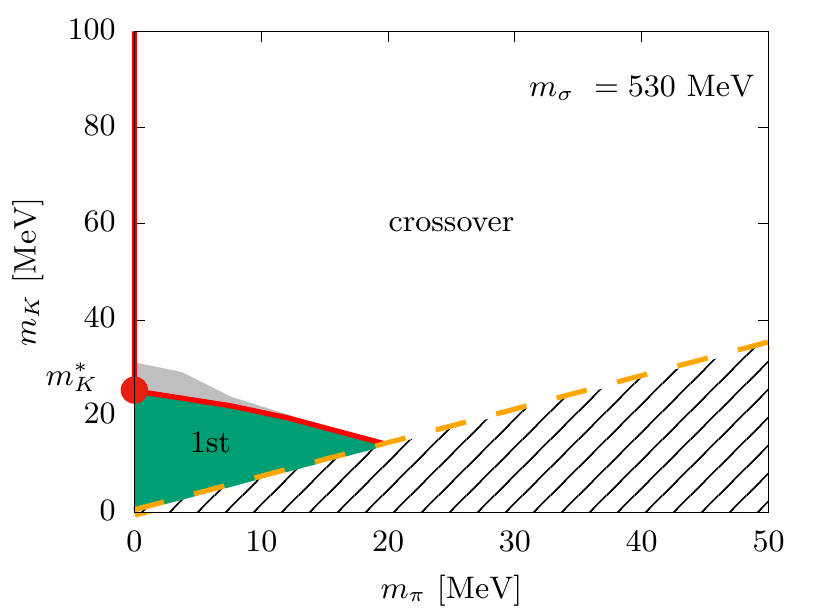}
  \caption{Chiral phase structure in the ($m_{\pi}$,$m_K$)-plane for
    $\mu=0$ including the axial anomaly obtained with the
    FRG. First-order and crossover regions are separated by a chiral
    critical line (solid red line). The red dot denotes a tricritical
    point at ($m^{*}_{\pi},m^{*}_K$) $\sim$ ($0, 23$) MeV.  The dashed
    region is excluded since it corresponds to $j_s < 0$. The gray region around the tricritical
    point gives an estimate of the IR flow sensitivity of the critical line (see text for
    details). }
  \label{fig:Columbia}
\end{figure}

In Fig.\ \ref{fig:Columbia} the Columbia plot for $\mu=0$ with vacuum
and thermal quark and meson fluctuations within the FRG framework is
presented.  On the $m_K$-axis a second-order chiral phase transition
line is found that terminates at the tricritical point
$m^{*}_K \approx 23$ MeV, which is significantly smaller than the
physical mass and the eMFA result. For smaller kaon masses surrounding
the chiral limit a small first-order region is formed that is limited
by a second-order line in agreement with the predictions of
\cite{Pisarski:1983ms}.  The physical mass point is far outside the
plot range and lies deep within the crossover region.  In comparison
to the eMFA findings, the first-order region around the chiral limit
shrinks drastically upon the additional inclusion of mesonic
fluctuations.

It is important to stop the RG flow as deep in the IR as possible
towards the chiral limit. The reason is that the lowest IR mass scale
in the system determines when the RG flows freeze out and the mesons
in the pseudoscalar octet become very light close to the chiral
limit. Integrating out fluctuations at very small scales,
$k \lesssim m_{\pi}/2$, is numerically expensive due to the convexity
of the effective potential where the flow in the flat potential region
approaches a singularity \cite{Litim:2006nn, Pawlowski:2014zaa}. We
used $k_\text{IR} = 70$ MeV throughout this work. In particular, the
location of the chiral critical line close to the tricritical point is
sensitive to IR-cutoff effects. The gray shaded area in
Fig.~\ref{fig:Columbia} gives a rough estimate for the sensitivity of
the critical line on $k_\text{IR}$. The upper boundary of this area
was extracted at $k_\text{IR}=100$ MeV while the lower boundary (red
line) denotes the critical line at $k_\text{IR}=70$ MeV. Between
$k=100$ MeV and $k=70$ MeV the tricritical point moves towards lower
kaon mass by $\Delta m_K\sim 6$ MeV while the location of the critical
line is already frozen for $k\lesssim 100\,\text{MeV}$ and
$m_\pi\gtrsim 10\,\text{MeV}$. We therefore expect only very minor
changes for even smaller $k_\text{IR}$ so that our main finding, the
surprisingly small first-order region around the chiral limit, will
not be altered. Scaling properties at the critical lines/points and a
precise determination of $m^{*}_{K}$ are not in the scope of the
present work.

\begin{figure*}
  \subfigure[$\;c_A\neq
  0$]{\includegraphics[width=\twofigs]{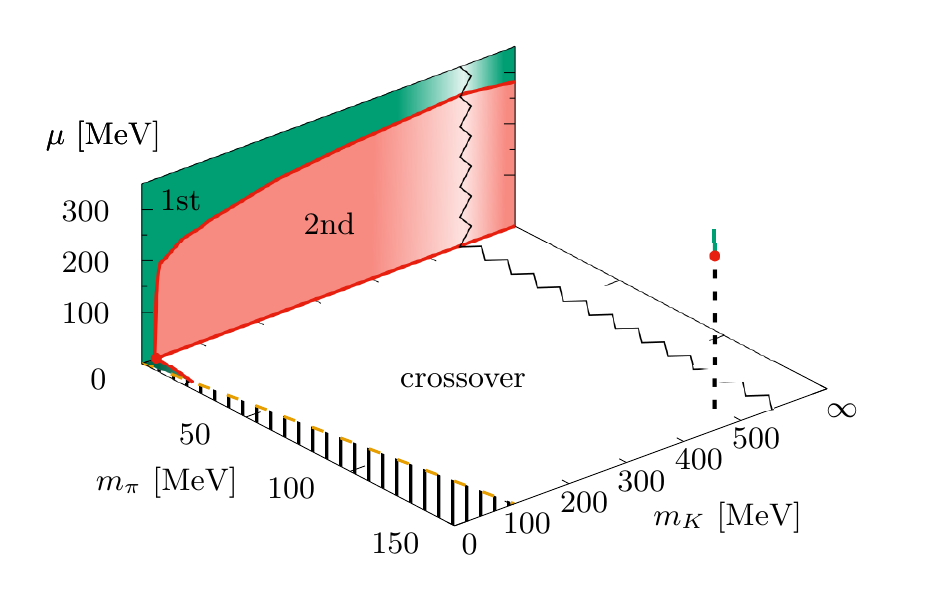}}
  \subfigure[$\;c_A=0$]{\includegraphics[width=\twofigs]{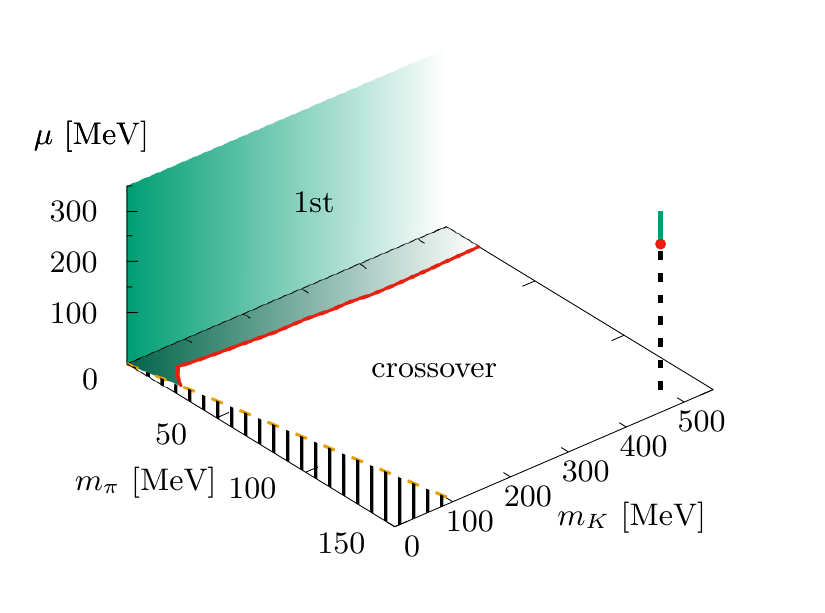}}
  \caption{Chiral phase structure in the ($m_{\pi} , m_K$)- and
    ($\mu, m_K$)-plane with the FRG (similar to
    Fig.~\ref{fig:columbiaPlotEMFA}). With $\ua$-symmetry breaking
    (left panel) and for $m_\sigma=530$ MeV a small first-order region
    is obtained around the chiral limit corner. Without the $\ua$
    breaking and for $m_\sigma=500$ MeV (right panel) the first-order
    region is extended along the $m_K$-axis. The vertical line
    indicates the physical mass point with a critical point (red
    point), crossover (black dashed line) and first-order transition
    (green solid line).}
  \label{fig:columbiaPlot3D}
\end{figure*}

The Columbia plots obtained with the FRG for vanishing as well as
finite $\mu$ with and without the axial anomaly are shown in
\Fig{fig:columbiaPlot3D}. With the axial anomaly (left panel in the
figure) the first-order region at $\mu=0$ around the chiral limit is
significantly smaller than the corresponding region in eMFA.  To be
more precise: towards the $SU(3)$-symmetric chiral limit, i.e. where
the light and the strange quark masses coincide and consequently
$m_\pi = m_K$, the boundary of the first-order region at $\mu=0$ is
located at a critical pion mass $m_\pi^c \sim 17$ MeV compared to
$m_\pi^c \sim 86$ MeV in the eMFA and $m_\pi^c \sim 150$ MeV in the
standard MFA. Thus, the conclusion is that vacuum quark as well as
meson fluctuations significantly reduce the size of the first-order
region in the Columbia plot around the chiral limit at $\mu=0$.

\begin{figure}[b] \centering
  \includegraphics[width=0.49\textwidth]{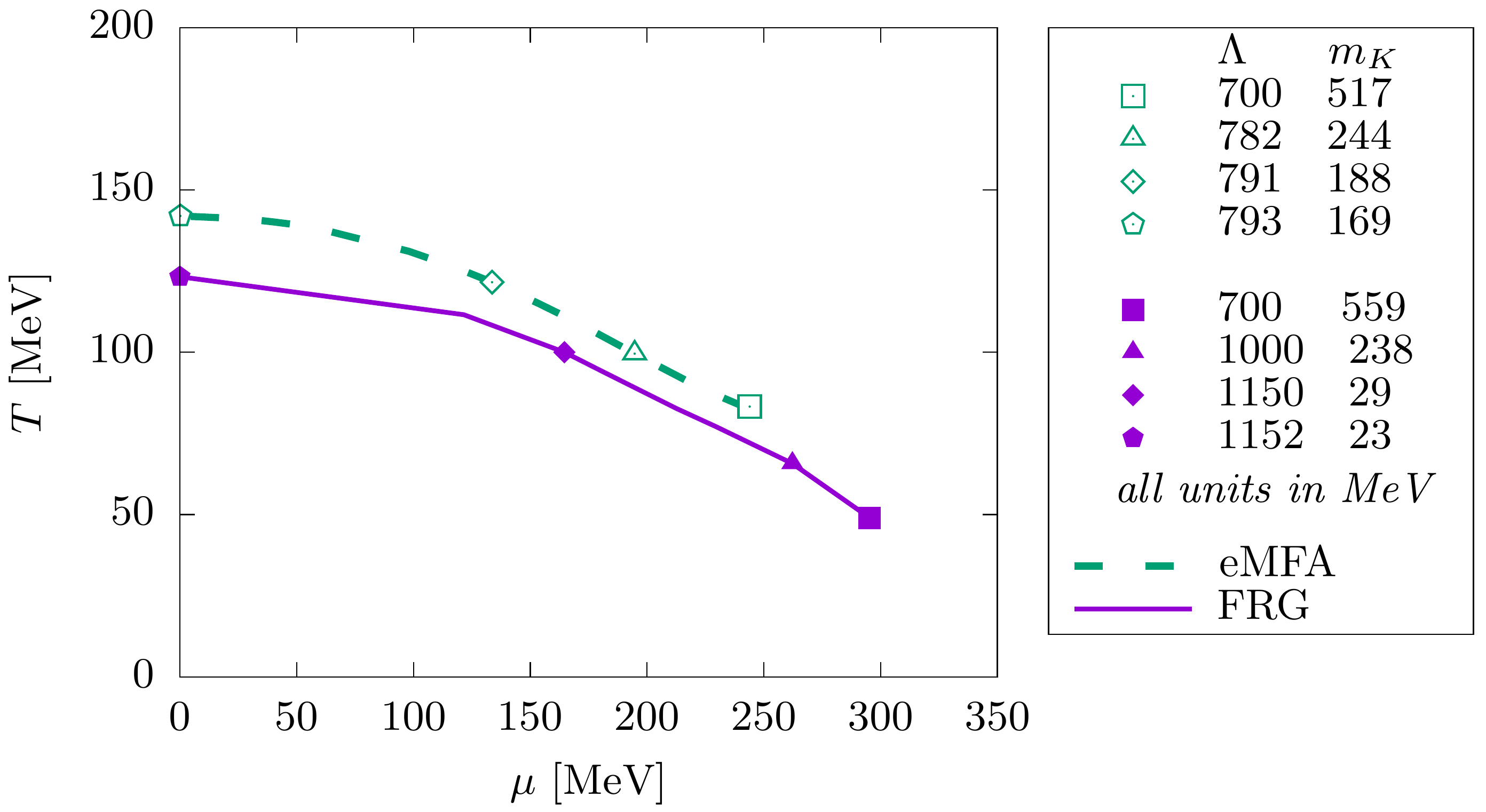}
  \caption{Kaon mass dependency of the CEP in the ($T,\mu$)-plane in
    the light chiral limit with the axial anomaly (dashed green open
    symbols: eMFA, solid purple symbols: FRG).}
  \label{fig:ceps} \end{figure}

The second-order chiral critical line extends from the tricritical
point at $m_{K}^{*} \approx 23$ MeV along the $m_K$-axis to the
two-flavor chiral limit. For $\mu>0$ the critical line steeply rises
to $\mu \approx 200$ MeV around $m_K \approx 50$ MeV and then slowly
grows to $\mu \approx 295$ MeV at physical kaon masses.  Similar to
the eMFA case the chiral critical surface has a positive slope such
that the size of the first-order region around the chiral limit
increases with increasing $\mu$. Also shown in the left panel of
\Fig{fig:columbiaPlot3D} is the extrapolation to the two-flavor chiral
limit. An explicit FRG calculation with a chiral
$SU(2)\otimes SU(2)$-symmetric quark-meson truncation with the same
sigma meson mass and $f_{\pi} = 92.4$ MeV yields a critical
$\mu_c = 281$ MeV which is also shown in the left panel of the
figure. However, this comparison should be viewed with some caution
since in a chiral $SU(2)\otimes SU(2)$-symmetric setup the axial
$\ua$-symmetry is maximally broken while in our $N_f = 2+1$ setup the
$\ua$-symmetry breaking, albeit large, is still finite.

Below the chiral critical line the transition for $m_{\pi}=0$ is
always of second-order as in the eMFA case but the area is clearly
larger in the FRG computation which can be attributed to the mesonic
fluctuations. In both calculations the tricritical point and the
two-flavor critical point are continuously connected by the chiral
critical line.

Our result without explicit $\ua$-symmetry breaking is shown in the
right figure of \Fig{fig:columbiaPlot3D}. The transition is always
first-order in the light chiral limit at $\mu=0$, independent of
$m_K$. This is in particular true for the chiral limits in the three-
and two-flavor cases, in agreement with \cite{Pisarski:1983ms}. Hence,
while we concluded in the last section that fermionic vacuum
fluctuations induce a second-order transition in the light chiral
limit in absence of the axial anomaly, we observe here that additional
meson fluctuations induce a first-order transition again. In the light
chiral limit the transition is always of first order, also at finite
$\mu$. We find a chiral critical line at vanishing density at
$m_\pi\sim 20\,\text{MeV}$ for all $m_K$ except very small ones, where
we see a slight bending towards larger $m_\pi$.

In \Fig{fig:ceps} the $m_K$-dependency of the CEP in the
($T, \mu$)-plane is shown for $c_A\neq 0$ in the light chiral limit.
Dashed green open symbols correspond to eMFA and solid purple symbols
are the FRG results.  Starting from the tricritical point at $\mu=0$
which has the highest critical temperature, the CEP moves towards
larger $\mu$ and smaller $T$ with increasing kaon mass.  The
$m_K$-dependency of the CEP is similar to the movement of the CEP when
the sigma meson mass is increased as seen in \cite{Schaefer:2008hk}.

We summarize our results in the chiral limits and compare with the
available literature in Tabs.~\ref{tab:supertab} and
\ref{tab:latcomptab}. The obtained order of the chiral transition in
different chiral limits with and without the axial anomaly is
confronted to other works in \Tab{tab:supertab}. The important
influence of quantum fluctuations and the axial anomaly is evident.

In \Tab{tab:latcomptab} the critical pion mass $m_\pi^c$ at $\mu=0$
and $c_A\neq 0$ for three degenerate quark flavors is compared to
available lattice results. Even recent lattice results show enormous
discrepancies depending on the implementation of fermions. The general
trend is that the closer one approaches the continuum limit, by
decreasing the lattice spacing or by improving the action, the lighter
$m_\pi^c$ becomes. It was argued in \cite{deForcrand:2017cgb} that
converged lattice results for the critical pion mass are expected to
be very small. Our continuum results are certainly in favor of a very
small critical $m_\pi^c$.

\begin{table*}
  \renewcommand{\arraystretch}{1.2}
  \begin{tabular}{ l | a b | a b | a b | a b}
\hline \hline
                                           & \multicolumn{2}{c|}{light chiral limit, $N_f = 3$}    & \multicolumn{2}{c|}{strange chiral limit, $N_f = 3$} & 
\multicolumn{2}{c|}{chiral limit, $N_f = 3$}            & \multicolumn{2}{c}{chiral limit, $N_f = 2$}                                                                                                                        
\\
                                           & $c_A>0$                                               & $c_A=0$                                              & $c_A>0$                                                 
& $c_A=0$                                                & $c_A>0$                  & $c_A=0$                  & $c_A>0$                  & $c_A=0$                  \\
\hline          
$\epsilon$-exp. LSM \cite{Pisarski:1983ms} & \sf{X}                                                & \sf{X}                                               & \sf{X}                                                  
& \sf{X}                                                 & \textbf{1$^{\text{st}}$} & \textbf{1$^{\text{st}}$} & \textbf{2$^{\text{nd}}$} & \textbf{1$^{\text{st}}
$} \\
CJT LSM \cite{Lenaghan:2000kr}             & \textbf{2$^{\text{nd}}$} $\forall\; m_K$              & \textbf{1$^{\text{st}}$} $\forall\; m_K$             
& \textbf{crossover}                                      & \textbf{1$^{\text{st}}$} $\forall\; m_l \lesssim 12$   & \textbf{2$^{\text{nd}}$} & 
\textbf{1$^{\text{st}}$} & \textbf{2$^{\text{nd}}$} & \textbf{1$^{\text{st}}$} \\
MFA QM \cite{Schaefer:2008hk}              & \textbf{1$^{\text{st}}$} $\forall\; m_K$              & \textbf{1$^{\text{st}}$} $\forall\; m_K$             
& \textbf{1$^{\text{st}}$} $\forall\; m_\pi \lesssim 177$ & \textbf{1$^{\text{st}}$} $\forall\; m_\pi \lesssim 50$ & \textbf{1$^{\text{st}}$} & 
\textbf{1$^{\text{st}}$} & \textbf{1$^{\text{st}}$} & \textbf{1$^{\text{st}}$} \\
eMFA QM                                    & \textbf{1$^{\text{st}}$} $\forall\; m_K \lesssim 169$ & \textbf{2$^{\text{nd}}$} $\forall\; m_K$             & 
\textbf{1$^{\text{st}}$} $\forall\; m_\pi \lesssim 110$ & \textbf{crossover}                                     & \textbf{1$^{\text{st}}$} & 
\textbf{2$^{\text{nd}}$} & \textbf{2$^{\text{nd}}$} & \textbf{2$^{\text{nd}}$} \\
FRG QM                                     & \textbf{1$^{\text{st}}$} $\forall\; m_K \lesssim 23$  & \textbf{1$^{\text{st}}$} $\forall\; m_K$             & 
\textbf{1$^{\text{st}}$} $\forall\; m_\pi \lesssim 20$  & \textbf{1$^{\text{st}}$} $\forall\; m_\pi \lesssim 25$ & \textbf{1$^{\text{st}}$} & 
\textbf{1$^{\text{st}}$} & \textbf{2$^{\text{nd}}$} & \textbf{1$^{\text{st}}$} \\
\hline \hline
\end{tabular}
\caption{Chiral phase transition order in different chiral limits with
  and without the axial anomaly from various methods (masses in MeV
  units). No predictions are labeled as $\sf{X}$. The transition is
  always of second-order above a given upper bound on $m_K$ in the
  light chiral limit and similarly a crossover in the strange chiral
  limit when an upper bound on $m_{\pi}$ is given.  In the CJK LSM
  \cite{Lenaghan:2000kr} the Columbia plot was explored in the
  ($m_l,m_s$)-plane and the results with the lightest available sigma
  mass (600 MeV) has been chosen. }
  \label{tab:supertab}
\end{table*}

\begin{table}[b]
  \renewcommand{\arraystretch}{1.2}
  \begin{tabular}{ c | c | c}
    \hline \hline
    method                                               & $m_\pi^c$ [MeV]   & Year \\
    \hline
    \rowcolor{columcol!7}[1.\tabcolsep]
    standard staggered, $N_t = 4$ \cite{Karsch:2001nf}   & $\sim 290$        & 2001 \\
    \rowcolor{columcol!20}[1.\tabcolsep]
    p4 staggered, $N_t=4$ \cite{Karsch:2003va}           & $\sim 67$         & 2004 \\
    \rowcolor{columcol!7}[1.\tabcolsep]
    standard staggered, $N_t=6$ \cite{deForcrand:2007rq} & $\sim 150$        & 2007 \\
    \rowcolor{columcol!20}[1.\tabcolsep]
    stout staggered, $N_t=6$ \cite{Varnhorst:2015lea}    & could be zero     & 2014 \\
    \rowcolor{columcol!7}[1.\tabcolsep]
    Wilson-clover, $N_t=6,8$ \cite{Jin:2014hea}          & $\sim 300$        & 2014 \\
    \rowcolor{columcol!20}[1.\tabcolsep]
    HISQ staggered, $N_t=6$ \cite{Bazavov:2017xul}       & $\lesssim 50$ & 2017 \\
    \rowcolor{columcol!7}[1.\tabcolsep]
    Wilson-clover, $N_t=8,10$ \cite{Jin:2017jjp}         & $\lesssim 170$    & 2017 \\
    \rowcolor{columcol!20}[1.\tabcolsep]
    MFA QM \cite{Schaefer:2008hk}                        & $\sim 150$        & 2008 \\
    \rowcolor{columcol!7}[1.\tabcolsep]
    eMFA QM                                              & $\sim 86$         & 2017 \\
    \rowcolor{columcol!20}[1.\tabcolsep]
    FRG QM                                               & $\sim 17$         & 2017 \\
    \hline \hline
  \end{tabular} 
  \caption{Critical pion mass $m_\pi^c$ 
for three degenerate quark flavors at
    $\mu=0$ of continuum QM model studies in
    comparison to various lattice QCD results (adapted from
    \cite{deForcrand:2017cgb}).  The lattice results are labeled with
    the corresponding fermion implementations and the number of time
    slices $N_t$.}
  \label{tab:latcomptab}
\end{table}

\section{Summary}\label{sec:sum}

We have studied the mass sensitivity of the chiral phase transition,
the left side in the Columbia plot, within a low-energy effective
description of QCD. To fully account for non-perturbative corrections
to the effective potential caused by quark and meson fluctuations, we
applied the functional renormalization group. In addition, we also
performed an extended mean-field computation within our framework in
order to discern between effects from different fluctuation sources.
This allowed us to systematically assess the impact of vacuum and
thermal fluctuations of quarks and mesons.

In the presence of the axial $\ua$ anomaly in the light chiral limit,
fermionic vacuum fluctuations induce a second-order phase transition
for larger kaon masses at vanishing density. A rather
extensive first-order region around the three-flavor chiral limit
emerges. It is separated from a crossover region by a chiral critical
line. Meson fluctuations significantly reduce the size of the
first-order region. In both the eMFA and the FRG computations the
tricritical point at $\mu=0$ and the two-flavor CEP at finite $\mu$
are continuously connected by a chiral critical line. The critical
surface bends away from the origin, i.e., the first-order region
increases for increasing $\mu$. The physical point always lies deep
within the crossover region at vanishing density. Consistent with the
curvature of the chiral critical surface a CEP
is found at rather large $\mu$.

For the $\ua$-symmetric system, we have shown that fermionic vacuum
fluctuations, on top of thermal fluctuations, induce a second-order
phase transition in the light chiral limit. As a consequence, there is
no first-order region in the Columbia plot. Meson fluctuations then
turn this second-order transition in the light chiral limit into first
order and move the chiral critical line to finite $m_{\pi}$
values. There is no CEP for any $\mu$ for very small pion masses since
the transition is always first-order. Thus, the effects of quantum
fluctuations on the order of the phase transition drastically depend
on the fate of the axial anomaly.

We have also demonstrated the crucial importance of a proper
definition of the initial action of the employed low-energy effective
model. Simply varying the current quark masses at the initial
scale of the effective theory results in the loss of spontaneous
chiral symmetry breaking in the chiral limit. This contradicts our
current knowledge about gauge theories with light fermion flavors. We
argued that the variation of the explicit chiral symmetry breaking
parameters leads to sizable modifications of the initial correlation
length and therefore concluded that the loss of spontaneous symmetry
breaking can indeed be attributed to an improper choice for the
initial action. We proposed the \textit{fixed} \textit{$f_\pi$-scheme}
for fixing the initial action for arbitrary values of the current
quark masses. Its motivation is drawn from $\chi$PT results on the
quark mass and flavor dependence of the decay constants. It implies
that the initial action has to be modified when the explicit chiral
symmetry breaking parameters are varied.

Our results outline the crucial role that fluctuations play for the
order of the chiral phase transition. We were able to identify how
different sources of quantum fluctuations affect the phase
transition. A major drawback of the present analysis is the absence of
gluon fluctuations. This can conveniently be taken into account with
the FRG by using dynamical hadronization along the lines of, e.g.,
\cite{Braun:2014ata}. The advantage of this approach is that the RG
flow can be started in the perturbative regime and the low-energy
physics is uniquely fixed by the parameters of microscopic QCD. In
addition to the access to the influence of gauge degrees of freedom on
the phase transition, it also cures the problem of a heuristic
determination of the initial effective action. Another drawback is the
implementation of the axial anomaly via a constant coupling to the 't
Hooft determinant. In a more realistic scenario, the anomaly coupling
$c_A$ depends on the medium parameters such as $T$ and $\mu$ as well
as on the RG scale $k$.

\subsection*{Acknowledgments}

We thank Konstantin Otto for providing us with some two-flavor data.
This work was supported by the FWF grant P24780-N27, the BMBF grants
05P15RGFCA and 05P12VHCTG, and by HIC for FAIR within the LOEWE
program of the State of Hesse. F.R.~also acknowledges funding by the
Deutsche Forschungsgemeinschaft (DFG) through grant RE 4174/1-1.\\

\appendix

\section{Numerical Setup}\label{sec:num}

The flow equation \eq{eq:flowEquation} is a highly non-linear partial
differential equation. We reduce this equation to an ordinary
differential equation by discretizing the effective potential on a
two-dimensional grid in the variables
\begin{align}
  x=\sigma_l^2\ ,\qquad y=2\sigma_s^2-\sigma_l^2\ .
\end{align}
The first and second derivatives of the potential with respect to the
fields, which are needed for the meson masses, are found by
interpolating the grid points with cubic splines including the
appropriate derivatives. The resulting coupled ordinary differential
equation for the potential at each grid point can be solved with any
time stepping algorithm. After the evolution towards the infrared to
some $k_\text{IR}$ we recover the full effective potential with
spline interpolation and evaluate the minimum of the IR potential
which yields the condensates $\sigma_l$ and $\sigma_s$ as well as the
meson masses.

Alternatively, one can solve the flow equation by a Taylor expansion
of the effective potential in powers of the fields around some
expansion point. For each coupling a coupled differential equation can
be derived representing the corresponding beta-function. In principle,
it is to some extend also possible to capture a first-order transition
with a local Taylor-expansion on a fixed background field
configuration, but the grid method is more suitable to compute the global
effective potential structure \cite{Pawlowski:2014zaa,
  Rennecke:2016tkm, Nakano:2009ps}. This is of more importance in
particular close to the chiral limit where the effective potential
becomes flat due to convexity.

The grid setup was tuned such that results converge for grid sizes of
the order of $N_x\times N_y=40\times 25$.  Since the symmetry breaking
parameters $c_l$ and $c_s$ are scale-independent and furthermore do
not enter in the meson masses, the symmetry breaking terms
$-c_l\sigma_l-c_s\sigma_s$ can be added to the evolved infrared
potential in the chiral limit.  Note that a similar treatment for the
$\ua$-symmetry breaking term is not possible since the 't Hooft
determinant is cubic in the meson fields and therefore enters in the
meson masses.

\bibliography{columbia.bib}

\end{document}